\documentclass[prl,twocolumn,showpacs,preprintnumbers,nofootinbib,amsmath,amssymb,nobalancelastpage]{revtex4}
\usepackage{graphicx}
\usepackage{bm}
\usepackage{dcolumn}
\usepackage{amsmath}
\usepackage{amssymb}
\usepackage{color}

\usepackage{hyperref}

\usepackage{aas_macros}

\newcommand{\GeV}{\mathrm{GeV}}

\newcommand{\sA}{\sigma}
\newcommand{\sigv}{\langle \sA v \rangle}

\newcommand{\degr}{^\circ}

\newcommand{\PSF}{\mathrm{PSF}}

\newcommand{\CDF}{\mathrm{CDF}}

\newcommand{\pglobal}{{p_{\rm global}}}

\begin{document}

\title{Indication of Gamma-ray Emission from the Newly Discovered Dwarf Galaxy Reticulum~II}
\author{Alex Geringer-Sameth}
\email{alexgs@cmu.edu}
\affiliation{McWilliams Center for Cosmology, Department of Physics, Carnegie Mellon University, Pittsburgh, PA 15213, USA}

\author{Matthew G. Walker}
\email{mgwalker@andrew.cmu.edu}
\affiliation{McWilliams Center for Cosmology, Department of Physics, Carnegie Mellon University, Pittsburgh, PA 15213, USA}

\author{Savvas M. Koushiappas}
\email{koushiappas@brown.edu}
\affiliation{Department of Physics, Brown University,  Providence, RI 02912, USA}

\author{Sergey E. Koposov}
\affiliation{Institute of Astronomy, University of Cambridge, Cambridge, CB3 0HA, UK}

\author{Vasily Belokurov}
\affiliation{Institute of Astronomy, University of Cambridge, Cambridge, CB3 0HA, UK}

\author{Gabriel Torrealba}
\affiliation{Institute of Astronomy, University of Cambridge, Cambridge, CB3 0HA, UK}

\author{N. Wyn Evans}
\affiliation{Institute of Astronomy, University of Cambridge, Cambridge, CB3 0HA, UK}

\date{\today}

\begin{abstract}
We present a search for $\gamma$-ray emission from the direction of the newly discovered dwarf galaxy Reticulum~II.  Using Fermi-LAT data, we detect a signal that exceeds expected backgrounds between $\sim 2-10$ GeV and is consistent with annihilation of dark matter for particle masses less than a few $\times\, 10^2\,\GeV$.  Modeling the background as a Poisson process based on Fermi-LAT diffuse models, and taking into account trials factors, we detect emission with $p$-value less than $9.8\times 10^{-5}$ ($>3.7\sigma$).  An alternative, model-independent treatment of background reduces the significance, raising the $p$-value to $9.7\times 10^{-3}$ ($2.3 \sigma$).  Even in this case, however, Reticulum~II has the most significant $\gamma$-ray signal of any known dwarf galaxy.  If Reticulum~II has a dark matter halo that is similar to those inferred for other nearby dwarfs, the signal is consistent with the $s$-wave relic abundance cross section for annihilation. 

\end{abstract}


\maketitle
Dark matter's non-gravitational interactions have profound implications for particle physics beyond the Standard Model, motivating searches for high-energy photons produced via annihilation.  The search for $\gamma$-rays in dwarf galaxies~\cite[e.g.][]{1990Natur.346...39L,2004PhRvD..70b3512B,2006PhRvD..73f3510B,  2007PhRvD..75b3513C,2006JCAP...03..003P,2007PhRvD..75h3526S,2010JCAP...01..031S, 2008ApJ...678..614S,2008ApJ...678..594W,2011APh....34..608H,2011PhRvL.107x1303G,2011PhRvL.107x1302A,2012APh....37...26M,2012PhRvD..86f3521B,2013arXiv1309.4780H,2013PhRvD..88h2002A,2013JCAP...03..018S,2013ApJ...773...61S,2014arXiv1409.1572C,2014JCAP...02..008A} provides an alternative to searches in regions that enjoy superior statistics but suffer from complicated backgrounds (e.g. the Galactic center~\cite{2011PhLB..697..412H,2011PhLB..705..165B,2011PhRvD..84l3005H,2012PhRvD..86h3511A,2013PhRvD..87l9902A,2013PhRvD..88h3521G,2014arXiv1402.4090A,2014arXiv1402.6703D,2014arXiv1406.6948Z,2014arXiv1409.0042C,2014arXiv1410.6168A}).  The observed stellar kinematics of dwarf galaxies imply gravitational potentials dominated by dark matter ~\cite{1998ARA&A..36..435M,2013pss5.book.1039W,2013PhR...531....1S,2013NewAR..57...52B}.  Many of these objects are nearby, are located at high galactic latitudes far from complicated emission regions, and possess no known astrophysical $\gamma$-ray sources.  Previous studies of dwarf galaxies have found no significant $\gamma$-ray emission, setting strong limits on the cross section for dark matter annihilation~\cite{2010ApJ...712..147A,2010JCAP...01..031S,2010PhRvD..82l3503E,2011PhRvL.107x1303G,2011PhRvL.107x1302A,2012APh....37...26M,2012PhRvD..86b1302G,2012PhRvD..86f3521B,2012PhRvD..86b3528C,2013JCAP...03..018S,2014PhRvD..89d2001A,2015PhRvD..91h3535G}

Using photometric data from the Dark Energy Survey (DES)~\cite{2005astro.ph.10346T}, \citet{2015arXiv150302079K} and \citet{2015arXiv150302584T} have recently announced the discovery of several low-luminosity Milky Way satellites in the Southern sky. 
\citet{2015arXiv150302079K} report 9 new objects. One of these, Reticulum~II (RetII), at a distance of 30~kpc, is the nearest dwarf galaxy after Segue 1 (Seg1, 23~kpc) and Sagittarius (24~kpc). RetII is $\sim3$ times more luminous than Seg1, suggesting that its dark matter halo may be more massive than Seg1's and making RetII an attractive place to search for a dark matter annihilation signal. 

Reticulum~II occupies a near-ideal location for $\gamma$-ray analysis: $49.7\degr$ below the Galactic plane and far from known $\gamma$-ray emitting sources (the closest source in the 3rd Fermi Catalog~\cite{2015arXiv150102003T} is $2.9\degr$ away). At energies above 1~GeV the $\gamma$-ray point spread function is significantly less than $1\degr$, making source contamination unlikely. Of the nearby dwarfs, only Seg1 is further from known sources. The interstellar emission model provided by the Fermi collaboration shows that emission from diffuse processes is relatively uniform within $10\degr$ of RetII.

We use Fermi-LAT data~\cite{2009ApJ...697.1071A} collected between August 8, 2008 and February 6, 2015. Using the publicly available {\tt Fermi Science Tools} (\url{http://fermi.gsfc.nasa.gov/ssc/}) (version {\tt v9r33p0}), we extract {\tt Pass 7 Reprocessed SOURCE} class events within $10\degr$ of RetII using {\tt gtselect} with {\tt zmax=$100\degr$}, and find good time intervals with {\tt gtmktime} with filter {\tt DATA\_QUAL==1 \&\& LAT\_CONFIG==1} and {\tt roicut=no}. The PSF and exposure in the direction of RetII are found by running {\tt gtselect} with a radius of $0.5\degr$, {\tt gtmktime} with {\tt roicut=yes}, {\tt gtltcube} with default options, and {\tt gtpsf} (with 17 log-spaced energies between 133.3 MeV and 1.333 TeV, {\tt thetamax=$10\degr$}, and {\tt ntheta=500}). 

The search for annihilation is based on event weighting as discussed in \cite{2015PhRvD..91h3535G}. The search suffers minimal loss in sensitivity when including only events within $0.5\degr$ of a dwarf galaxy and with energies above 1~GeV (see Figs. 3--5 of~\cite{2015PhRvD..91h3535G}). We adhere to these criteria in this analysis and define a region of interest (ROI) as a region of radius $0.5\degr$ containing events between 1--300~GeV. Gamma-ray sources from the 3rd Fermi Catalog are assigned masks of at least $0.8\degr$ (the approximate PSF at 1~GeV).

\begin{figure}
\includegraphics{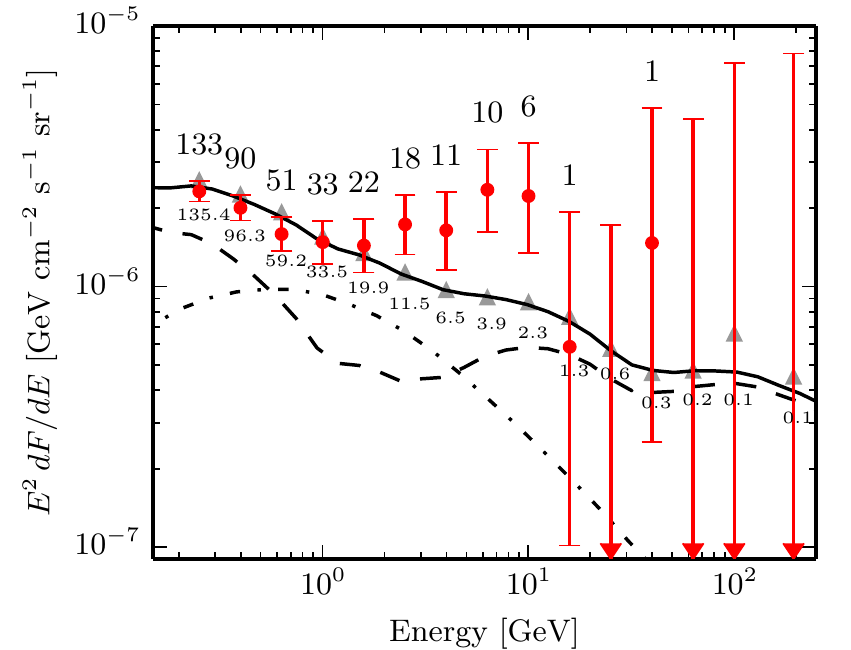}
\caption{\label{fig:Ret2spec}Energy spectrum of events detected within $0.5\degr$ of RetII (red points) with 68\% Poisson error bars. Two background estimates are shown: 1) the sum (solid black) of the Fermi Collaboration's models for isotropic (dashed)  and galactic diffuse (dot dash) emission at the location of RetII, and 2) the average intensity (gray triangles) within 3306 ROIs that lie within $10\degr$ of RetII and overlap neither known sources nor the ROI centered on RetII.  The number of events detected from RetII in each energy bin is shown above the error bar. The number expected from the Fermi background model is shown below the solid black curve.}
\end{figure}

Figure~\ref{fig:Ret2spec} shows the energy spectrum derived from an ROI centered on RetII (red points). For each energy bin, the differential flux $dF/dE$ is the number of events divided by the width of the energy bin, the instrument exposure, and the ROI's solid angle. Error bars indicate standard 68\% Poisson confidence intervals~\cite[e.g.][]{casella2002statistical} on the mean counts in each bin (5 bins per decade between 0.2~GeV and 300~GeV). The figure also shows two estimates of background.  First, the solid black line represents a two-component background model that is derived by the Fermi collaboration (\url{http://fermi.gsfc.nasa.gov/ssc/data/access/lat/BackgroundModels.html}).  It is the sum of the isotropic spectrum {\tt iso\_source\_v05.txt} (dashed black line) and the diffuse interstellar emission model {\tt gll\_iem\_v05\_rev1.fit} (dot dashed).  The latter is averaged over the $1\degr$ region surrounding RetII (we confirmed that the curve does not change for any choice of radius within $5\degr$).  Second, gray triangles indicate an empirical estimate of background, showing the average intensity within 3306 ROIs that fall within $10\degr$ of RetII and do not overlap with any source masks, the central ROI, or the boundary of the $10\degr$ region (see Fig.~\ref{fig:bgsamples}, right panel).  The two estimates of background show good agreement.  \textit{Between 2 GeV and 10 GeV, the spectrum from RetII clearly rises above the expected background.}

To derive a detection significance we employ the following method (see~\cite{2015PhRvD..91h3535G} for details). 
Each event in the ROI is assigned a weight $w(E,\theta)$ based on its energy $E$ and angular separation $\theta$ from the ROI center. The test statistic $T = \sum w(E_i, \theta_i)$ is the sum of the weights of all events in the ROI, with larger values of $T$ providing evidence of a signal. In this approach, the most powerful weight function for testing the background-only hypothesis is given by $w(E, \theta) = \log [ 1 + s(E,\theta) / b(E,\theta)]$, where $s(E,\theta)$ is the expected number (in a small $dE, d\theta$ range) of events due to dark matter annihilation for the alternative hypothesis (signal) and $b(E,\theta)$ is the expected number from all other sources (background).  

The expected signal depends on the dark matter particle properties (mass $M$, annihilation cross section $\sigv$), the dark matter content of the dwarf galaxy (parameterized here by the single quantity $J$~\cite[e.g.][]{0004-637X-801-2-74}), and the detector response (exposure $\epsilon$ and PSF):
\begin{equation}
\frac{s(E,\theta)}{dE d\theta} = \frac{\sigv J}{8 \pi M^2}  \frac{dN_f(E)}{dE} \times \epsilon(E) \PSF(\theta|E) 2\pi \sin(\theta).
\label{eqn:expectedsignal}
\end{equation}
For annihilation into a final state $f$, $dN_f/dE$ is the number of $\gamma$-rays produced (per interval $dE$) per annihilation.  We adopt the annihilation spectra of~\citet{2011JCAP...03..051C}, which include electroweak corrections~\cite{2011JCAP...03..019C}. Note that the unknown $J$ value is exactly degenerate with $\sigv$.  

We quantify the signal's significance by calculating its $p$-value: the probability that background could generate events with a total weight greater than that observed for the ROI centered on RetII. We also quote ``$\sigma$~values'', $\CDF^{-1}(1-p)$, using the standard normal CDF.

First we compute significance by modeling the background in the central ROI as an isotropic Poisson process.  This procedure is justified by RetII's location in a quiet region that is far from known sources and strong gradients (see Fig.~\ref{fig:bgsamples}, right panel).  Specifically, we assume that 1)  the number of background events within $0.5\degr$ of RetII is a Poisson variable, 2) background events are distributed isotropically, and 3) their energies are independent draws from a given spectrum.  Under these assumptions the test statistic is a compound Poisson variate whose PDF we can calculate for any weight function and any adopted background spectrum~\cite{2015PhRvD..91h3535G}.  There is no assumption that the PDF follows an asymptotic form such as $\chi^2$.

\begin{figure*}
\includegraphics[scale=1.]{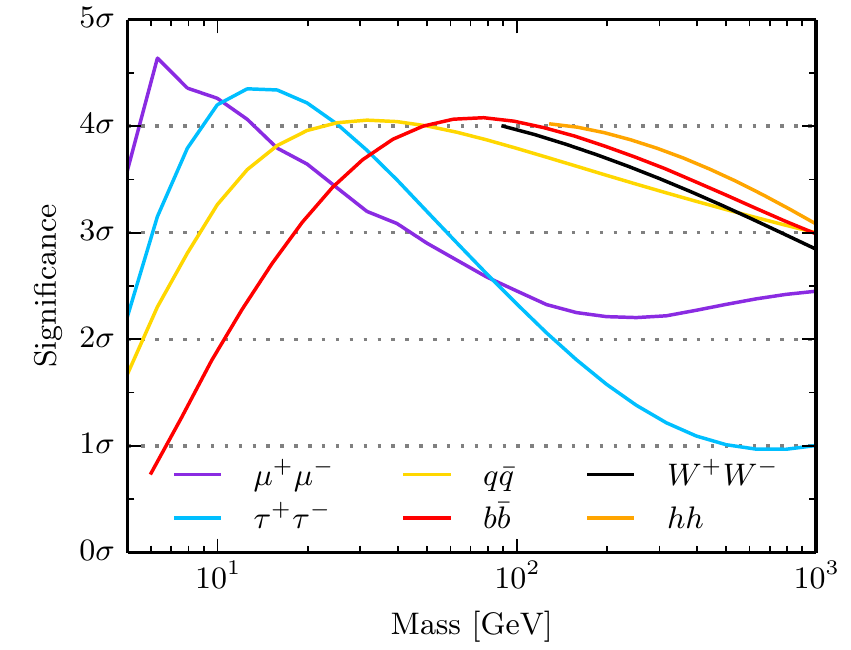}
\includegraphics[scale=1.]{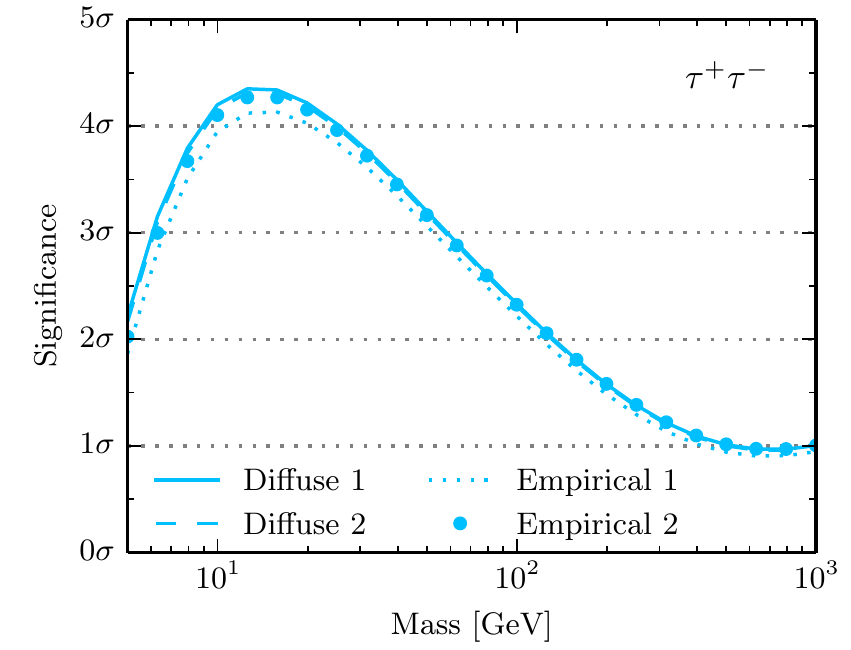}
\caption{\label{fig:bgpoisson}Significance of the $\gamma$-ray excess in the direction of Reticulum~II as a function of dark matter particle mass. {\it Left:} Curves correspond to the result of the search in various channels (i.e. using different ways of weighting events) using background model Diffuse 1. The curve for $e^+e^-$ is similar to $\mu^+\mu^-$,  $ZZ$ is similar to $W^+W^-$, and $q$ represents $u,d,c,s$ quarks and gluons. {\it Right:} Significance in the $\tau^+\tau^-$ channel for four different background models (see text).}
\end{figure*}

We consider four possible energy spectra for the background $b(E,\theta)$.  The first two are sums of the Fermi collaboration's isotropic and galactic-diffuse models, where the latter is averaged within either $1\degr$ or $2\degr$ of RetII.  We refer to these spectra as `Diffuse 1' (this is the same background model shown in Fig.~\ref{fig:Ret2spec}) and `Diffuse 2'.  The third is an empirically-derived spectrum (`Empirical 1') using events between $1\degr$ and $5\degr$ from RetII (excluding masked sources).  Below 10 GeV, this spectrum is a kernel density estimate, with each event replaced by a Gaussian with width $20\%$ of its energy.  Above 10 GeV we fit a power law with exponential cutoff.  Finally, we bin the same events (30 bins between 0.2 GeV and 1 TeV) in order to construct a fourth possible background spectrum (`Empirical 2'), where the intensity between bin centers is found by linear interpolation in $\log(\rm intensity)$. Figure~\ref{fig:bgpoisson} shows the significance of the detected $\gamma$-ray signal from RetII for various annihilation channels and for each background model. In every case, the significance peaks above $4\sigma$, with little dependence on choice of background spectrum.

However, it is important to consider a ``trials factor'' to account for the fact that we are searching for dark matter particles of any mass, i.e. conducting multiple hypothesis tests on the same data. As shown in Fig.~6 of~\cite{2015PhRvD..91h3535G}, the search is not particularly sensitive to the particle mass used in the weight function: $\sim3$ trial masses suffice if the true mass is between 10 GeV and 1 TeV for the $b\bar{b}$ and $\tau^+\tau^-$ channels. Nonetheless, we quantify the trials factor by simulating large numbers of ROIs under the Diffuse 1 model. A $p$-value is found at each trial mass and the minimum of these $p_m$ is recorded for each simulated ROI. The ``global'' $p$-value $\pglobal$ is the fraction of simulated ROIs with $p_m$ less than that observed in RetII.  Simulating $\sim 30$ million background ROIs, we find $\pglobal = 9.8\times 10^{-5}$ for $b\bar{b}$ and $\pglobal = 4.2 \times 10^{-5}$ for $\tau^+\tau^-$. Note that the trials factor may have a more significant effect for a lighter final state (e.g. electrons). 

Following~\cite{2011PhRvL.107x1303G,2012PhRvD..86b1302G,2015PhRvD..91h3535G}, we also consider an entirely different procedure for computing significance.  Under this second procedure, we construct the PDF of $T$ due to background by making a histogram of $T$ values for ROIs distributed over the region surrounding the dwarf.  This procedure is model-independent and automatically accounts for non-Poisson background processes (e.g. due to unresolved sources), an effect examined by several groups~\cite{2009JCAP...07..007L,2010PhRvD..82l3511B,2011PhRvL.107x1303G,2014PhRvD..89d2001A,2014arXiv1409.1572C,2015PhRvD..91h3535G,2014arXiv1412.6099L}.

\begin{figure*}
\includegraphics[scale=1.]{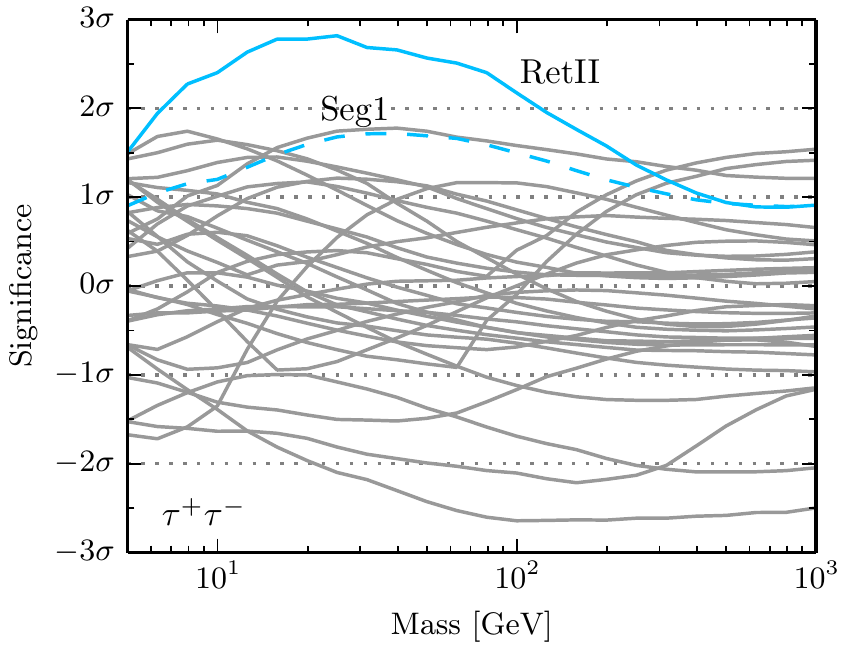}
\includegraphics[scale=1.]{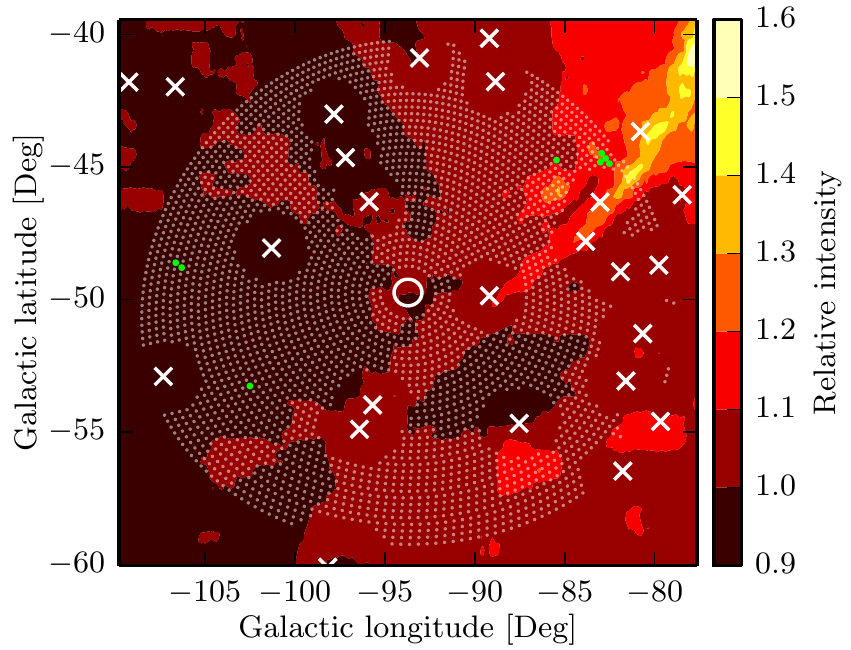}
\caption{\label{fig:bgsamples} {\it Left}: Significance of $\gamma$-ray detection for annihilation into $\tau^+\tau^-$ for various masses, calculated using the model-independent procedure of ~\cite{2015PhRvD..91h3535G}.  Solid and dashed blue lines correspond to RetII and Seg1 (another attractive nearby target). Gray curves correspond to the collection of dwarfs used in~\cite{2015PhRvD..91h3535G} as well as the 8 other newly discovered DES dwarfs. {\it Right:} The Fermi isotropic+diffuse model intensity near RetII. The color corresponds to intensity normalized to the value in the direction of RetII (at an energy of 8 GeV --- other energies are similar). A $0.5\degr$ ROI is shown at the center and the small dots show the centers of the ROIs used for the empirical background estimation. White $\times$'s mark the locations of known $\gamma$-ray sources.  Green circles are the ROIs which have a test statistic larger than that in the central ROI (when searching for a 25 GeV particle annihilating to $\tau^+\tau^-$).  }
\end{figure*}

The left-hand panel of Fig.~\ref{fig:bgsamples} shows the significance of RetII's signal as calculated following the model-independent procedure.  Compared with the Poisson-process model for background (see above), this procedure assigns less significance to RetII's $\gamma$-ray signal (in accord with~\cite{2015PhRvD..91h3535G,2014PhRvD..89d2001A,2014arXiv1409.1572C}).  For example, when searching for a 25 GeV particle annihilating to $\tau^+\tau^-$, eight of 3306 background ROIs have $T$-values larger than RetII's ($2.8\sigma$; other channels show similar reductions in significance).  

A trials factor for the model-independent approach is found by counting the number of background ROIs which have $T$ values among the top $n$ for at least one mass considered ($n$ is the rank of the central ROI at the most significant mass). For annihilation into $\tau^+\tau^-$, $n=9$ and there are 32 such ROIs, giving a global $p$-value of $32/3306 = 0.0097$ ($2.3\sigma$). The same global significance is found by computing what fraction of simulated Poisson background ROIs have a minimum $p$-value less than $8/3306$.

The application of this model-independent procedure to RetII reveals its fundamental limitation: a strong signal necessarily implies that very few background ROIs have $T$ larger than that of the object of interest. Thus, poor sampling of the large-$T$ tail prevents a robust calculation of significance for the RetII signal. For example, had we used a $5\degr$ background region instead of $10\degr$, {\em zero} background ROIs would have given a $T$ value larger than RetII, indicating that the significance calculation breaks down when there are not enough ``independent'' background regions. In any case, this procedure clearly identifies RetII's as the most tantalizing $\gamma$-ray signal from any known dwarf galaxy (left-hand panel of Fig.~\ref{fig:bgsamples}).

If the $\gamma$-ray signal is interpreted as dark matter annihilation, we perform a simple exploration of the allowed particle parameter space. As shown in~\cite{2015PhRvD..91h3535G}, for the two parameters $M$ and $\sigv$, the likelihood ratio is related to $T$: 
\begin{equation}
\log \frac{L(\text{data} \mid (M, \sigv)+\text{bg})}{L(\text{data} \mid \text{bg})} = T - \int_{E,\theta} s(E,\theta),
\end{equation}
where the integral is the expected number of events in the ROI due to dark matter annihilation. We denote the right-hand side as $\lambda(M,\sigv)$. 
Maximizing $\lambda(M,\sigv)$ yields the maximum likelihood estimate $\widehat{M},\widehat{\sigv}$.
The difference $2\lambda(\widehat{M},\widehat{\sigv}) - 2\lambda(M,\sigv)$ is distributed as a $\chi^2$ variable with 2 degrees of freedom~\cite{wilks1938} when $M,\sigv$ are the true values of the mass and cross section. Therefore, regions of $(M,\sigv)$ space where this difference is less than 2.3, 6.2, and 11.8 constitute 68.2\%, 95.4\%, and 99.7\% confidence regions. The $\chi^2$ behavior holds only for large sample sizes and it is not clear if that assumption is valid here. In particular, for annihilation into electrons or muons, where low masses are preferred, there are very few events above 1 GeV but below the dark matter mass.

\begin{figure}
\includegraphics[scale=1]{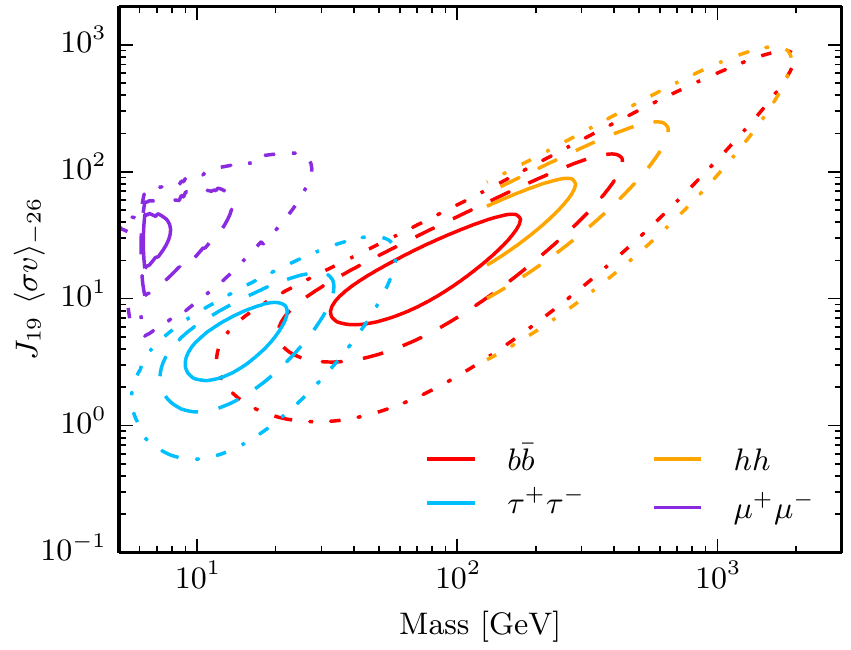}
\caption{\label{fig:mJsigv} An exploration of a dark matter interpretation of the observed $\gamma$-ray excess for four representative annihilation channels. $J = J_{19}\, 10^{19} {\mathrm{GeV^2cm^{-5}}}$ and $\sigv = \sigv_{-26}\, 10^{-26} {\mathrm{cm^3\sec^{-1}}}$. The data constrain only the product of $J \sigv$ since the dark matter content of Reticulum~II is currently unknown. Contours represent 68\%, 95\%, and 99.7\% confidence regions. Note that this figure does not quantify which annihilation channel is preferred by the data, i.e. which channel provides the best fit to the $\gamma$-ray spectrum.  }
\end{figure}

Figure~\ref{fig:mJsigv} shows the derived constraints on the product $J \sigv$ for a number of representative channels.  Although we cannot make a direct measurement of the cross section, the constraints on $J\sigv$, combined with independent knowledge of $\sigv$, allow us to make a {\em prediction} for the dark matter content of RetII which must hold if the $\gamma$-ray emission is due to annihilating dark matter. In the $\tau^+\tau^-$ channel, for example, dividing the maximum likelihood estimate of $J\sigv$ (Fig.~\ref{fig:mJsigv}) by the 95\% upper limit on $\sigv$ from~\cite{2015PhRvD..91h3535G} yields $\log_{10}J \gtrsim 19.6 \pm 0.3$, where the uncertainty reflects the 68\% confidence region shown in Fig.~\ref{fig:mJsigv}. For comparison, Seg1 has $\log_{10} J = 19.3 \pm 0.3$~\cite{0004-637X-801-2-74}.

While RetII's $\gamma$-ray signal is tantalizing, it would be premature to conclude it has a dark matter origin.  Among alternative explanations, perhaps the most mundane is the possibility that an extragalactic source lies in the same direction. Computing $T$ as a continuous function of sky position reveals the peak $T$-value to occur $0.083\degr$ from the optical center of RetII~\cite{2015arXiv150302079K}, an offset similar to typical localization errors for weak, high-energy sources in the 3rd Fermi Catalog~\cite{2015arXiv150102003T}. Thus the emission is consistent with originating from RetII's location. Searching the BZCAT~\cite{2009A&A...495..691M} and CRATES~\cite{2007ApJS..171...61H} catalogs reveals a CRATES quasar (J033553-543026) that is $0.46\degr$ from RetII. Further work must be done to determine whether this particular source contributes to the emission, though we note that flat spectrum radio quasars rarely have a spectral index less than 2~\cite{2011ApJ...743..171A,2015arXiv150106054A}. 
Other diagnostics, such as color-color diagrams, multiwavelength surveys, and variability searches, may eventually reveal the presence of active galaxies behind RetII. These must then be considered as possible $\gamma$-ray emitters. However, we emphasize that even without knowledge of specific background objects, the $p$-value derived from the background sampling procedure (Fig.~\ref{fig:bgsamples}) automatically accounts for the probability that a chance alignment is causing RetII's $\gamma$-ray signal.

There is also the possibility that $\gamma$-ray emission arises from within RetII, albeit through conventional processes. One of the much-discussed astrophysical explanations for the apparent Galactic Center excess is millisecond pulsars~\cite{2012PhRvD..86h3511A,2013PhRvD..88h3521G,2014JHEAp...3....1Y,2013MNRAS.436.2461M,2013PhRvD..88h3009H,2014arXiv1407.5625C,2014arXiv1407.5583C}. In the case of RetII, it is the high-energy behavior of the emission which disfavors a pulsar model, as millisecond pulsars exhibit an exponential cutoff at around 2.5 to 4~GeV~\cite{2013ApJS..208...17A,2013PhRvD..88h3521G,2014arXiv1409.0042C,2014arXiv1407.5583C,2014arXiv1412.2422M,2015JCAP...02..023P}. Alternatively, high-energy cosmic ray production could potentially arise in the vicinity of young massive stars. Upcoming photometric and spectroscopic analysis of RetII will check this possibility.

Thorough explorations of the diffuse background, the $\gamma$-ray events toward RetII, properties of RetII's dark matter halo, and any coincident sources will prove crucial to confirming or ruling out the dark matter interpretation. Fermi's upcoming Pass 8 data release~\cite{2013arXiv1303.3514A,2014AAS...22325603G} will improve every aspect of the instrument response, allowing for a more sensitive analysis of RetII and other known and as-yet-unknown Milky Way companions. Understanding the $\gamma$-ray emission, along with the analysis of RetII as a galaxy embedded in a dark matter halo, may provide a long-sought avenue for the characterization of dark matter particles.

We note that the Fermi collaboration has simultaneously performed an independent search for $\gamma$-ray emission and reports no significant excess from any dwarf galaxy, including RetII~\cite{2015arXiv150302632T,2015arXiv150302641F}.
Nevertheless, the strongest signal they find ($p=0.06$), for any annihilation channel and mass, corresponds to a 25~GeV particle annihilating into $\tau^+\tau^-$ in RetII (cf. our Fig.~\ref{fig:bgsamples}). 
The reason for any discrepancy with our result is unclear, as the Fermi analysis is based on unreleased data.

\begin{acknowledgments}
AGS gratefully acknowledges helpful discussions with Sukhdeep Singh. MGW is supported by NSF grants AST-1313045 and AST-1412999. SMK is supported by DOE DE-SC0010010, NSF PHYS-1417505, and NASA NNX13AO94G. The research leading to these results has received support from the European Research Council under the European Union's Seventh Framework Program (FP/2007-2013) ERC Grant Agreement no. 308024.

\end{acknowledgments}

\bibliography{manuscript}

\end{document}